\documentclass[preprint,showpacs,preprintnumbers,amsmath,amssymb]{revtex4}

\usepackage{graphicx}% Include figure files
\usepackage{dcolumn}% Align table columns on decimal point
\usepackage{bm}% bold math

\begin{document}

\preprint{}

\title{Moments of the Hilbert-Schmidt probability distributions over  determinants  of real two-qubit density matrices and of their partial transposes}
\author{Paul B. Slater}% 
\email{slater@kitp.ucsb.edu}
\affiliation{%
ISBER, University of California, Santa Barbara, CA 93106\\
}%
\date{\today}% It is always \today, today,

\begin{abstract}
The nonnegativity of the determinant of the partial transpose of a two-qubit ($4 \times 4$) density matrix ($\rho$) is both a necessary and sufficient condition for the separability of $\rho$. While the determinant of $\rho$ itself is restricted to the interval 
$[0,\frac{1}{256}]$, the determinant of the partial transpose 
($|\rho^{PT}|$)  can range over
$[-\frac{1}{16},\frac{1}{256}]$, with negative values corresponding to entangled states. We report here the exact values of the first nine moments of the probability distribution of 
$|\rho^{PT}|$ over this interval, with respect to the Hilbert-Schmidt (metric volume element) measure on the nine-dimensional convex set of real two-qubit density matrices. Rational functions 
$C_{2 j}(m)$, yielding the coefficients of the $2j$-th  
power of even polynomials occurring at intermediate steps in our derivation of the $m$-th moment, emerge. These functions  possess poles at finite series of consecutive half-integers 
($m=-\frac{3}{2},-\frac{1}{2},\ldots,\frac{2 j-1}{2}$), and certain 
(trivial) roots at finite series of consecutive natural numbers ($m=0, 1,\ldots$). Additionally, the (nontrivial)
dominant roots of  $C_{2 j}(m)$ approach the same half-integer values ($m=
\frac{2 j-1}{2}, \frac{2 j-3}{2},\ldots$), as $j$ increases. 
The first two moments (mean and variance) found--when employed in the 
one-sided Chebyshev inequality--give an upper bound
of $\frac{30397}{34749} \approx 0.874759$ on the separability probability 
 of real two-qubit density matrices. We are able to report general formulas for the $m$-th moment of
the Hilbert-Schmidt probability distribution of 
$|\rho|$ over $[0,\frac{1}{256}]$, in the real, complex and 
quaternionic two-qubit cases.
\end{abstract}


\maketitle
\section{Introduction}
One interesting, and seemingly not immediately obvious consequence of certain well-known results of  Peres and the Horodecki clan 
\cite{asher,michal} is that one only needs to evaluate the sign
of the determinant of the partial transpose of a two-qubit ($4 \times 4$) density matrix 
($\rho$) to assess the separability of $\rho$ 
\cite{sanpera5,ver2,augusiak,azuma}, rather than checking individually the signs of its four eigenvalues 
(since no more than one eigenvalue of $|\rho^{PT}|$ can be negative).
If one assigns a measure--we will here use the volume element of the Hilbert-Schmidt 
(Euclidean/flat) metric \cite{szHS} (cf. \cite{dunkl})--to the two-qubit density matrices, then, from the associated probability distribution over the determinant of the partial transpose, one should--in principle--be able to derive the specific and long-sought probability that a two-qubit density matrix is separable (cf. \cite{ZHSL,ye2,giraud1}). To attempt to fully characterize such a probability distribution of interest, we begin by computing its first several moments (sec.~\ref{sectranspose}). (It has been conjectured that "most of the information defining a compactly supported [probability distribution function] is usually contained in its first few moments" 
\cite{gavriliadis2,gavriliadis1,john}.) As a complementary exercise, we similarly analyze--but with considerably less  severe computational demands--the Hilbert-Schmidt probability distribution over the determinant $|\rho|$ itself (sec. \ref{secnotranspose}).
The results obtained allow us to construct a general formula (\ref{generalformula}) 
for the $m$-th moment of this distribution.

We will proceed in the framework of the
Bloore (or correlation coefficient) parameterization of the $4 \times 4$ density matrices \cite{bloore,joe,slaterPRA2} which allows us (in the generic real two-qubit case of immediate interest here) to work primarily in seven dimensions, rather than  the nine naively expected. Also, in our computations, we will further reparametrize three of the six correlations
\begin{equation}
z_{ij}=\frac{\rho_{ij}}{\sqrt{\rho_{ii} \rho_{jj}}}, \hspace{.1in} 1 \leq i <j \leq 4, \hspace{.2in} 
z_{ij} \in [-1,1]
\end{equation}
in terms of {\it partial} correlations \cite{joe}, allowing certain requisite integrations to be performed simply over six-dimensional hypercubes, rather than more complicated domains. (We had alternatively attempted to utilize the cylindrical algebraic decomposition 
\cite{cylindrical}
to define the integration limits (as indicated in 
\cite[sec. II]{slaterPRA2}) that specify the domain of feasible density matrices, directly within the Bloore-type framework, without  transforming to such partial correlations. However, we 
encountered certain apparently inconsistent/puzzling results obtained using Mathematica in this regard.)
\section{Hilbert-Schmidt probability distribution 
over $|\rho^{PT}|$} \label{sectranspose}
The computations of the $m$-th moment proceeds in two stages. In the first,
we perform an integration over the six-dimensional hypercube $[-1,1]^6$ of the $m$-th power of a (transformed) polynomial ($\tilde{P}$)--proportional to $|\rho^{PT}|$--in seven variables (\cite[eq. (7)]{slaterPRA2}). (The proportionality factor is $(\rho_{22} \rho_{33})^{2 m}$.) 
The free (unintegrated) variable is of the form
\begin{equation} \label{substitution}
\mu=\sqrt{\frac{\rho_{11} \rho_{44}}{\rho_{22} \rho_{33}}},
\end{equation}
where the $\rho_{ii}$'s are the diagonal entries of $\rho$. (In the related study \cite{slaterPRA2}, $\nu=\mu^2$ was used as a variable, and in 
\cite{advances}, $\xi=\log{\mu}$.) 
We have that (before the transformation to partial correlations, yielding $\tilde{P}$)
\begin{equation}
P=-z_{\text{il}}^2 \mu ^4+2 z_{\text{il}} \left(z_{\text{ij}}
   z_{\text{ik}}+z_{\text{jl}} z_{\text{kl}}\right) \mu ^3+2
   z_{\text{jk}} \left(z_{\text{ij}} z_{\text{jl}}+z_{\text{ik}}
   z_{\text{kl}}\right) \mu -z_{\text{jk}}^2 +
\end{equation}
\begin{displaymath}
\mu ^2 \left(\left(z_{\text{kl}}^2-1\right) z_{\text{ij}}^2-2
   \left(z_{\text{il}} z_{\text{jk}}+z_{\text{ik}} z_{\text{jl}}\right)
   z_{\text{kl}} z_{\text{ij}}+z_{\text{il}}^2
   z_{\text{jk}}^2-z_{\text{jl}}^2-z_{\text{kl}}^2-2 z_{\text{ik}}
   z_{\text{il}} z_{\text{jk}} z_{\text{jl}}+z_{\text{ik}}^2
   \left(z_{\text{jl}}^2-1\right)+1\right).
\end{displaymath}
The transformation of the three correlations $z_{il}, z_{ik},z_{jl}$ to partial correlations ($z_{ik,j},z_{jl,k},z_{il,jk}$)
takes the form \cite{joe}
\begin{equation}
z_{\text{il}}\to z_{\text{ij}} z_{\text{jk}}
   z_{\text{kl}}+\sqrt{z_{\text{ij}}^2-1} \sqrt{z_{\text{jk}}^2-1}
   z_{\text{ik},j} z_{\text{kl}}+z_{\text{ij}} \sqrt{z_{\text{jk}}^2-1}
   \sqrt{z_{\text{kl}}^2-1} z_{\text{jl},k} +
\end{equation}
\begin{displaymath}
\sqrt{z_{\text{ij}}^2-1} \sqrt{z_{\text{kl}}^2-1}
   \sqrt{z_{\text{ik},j}^2-1} \sqrt{z_{\text{jl},k}^2-1}
   z_{14,23}+\sqrt{z_{\text{ij}}^2-1} z_{\text{jk}}
   \sqrt{z_{\text{kl}}^2-1} z_{\text{ik},j} z_{\text{jl},k},
\end{displaymath}
\begin{displaymath}
z_{\text{ik}}\to z_{\text{ij}} z_{\text{jk}}+\sqrt{z_{\text{ij}}^2-1}
   \sqrt{z_{\text{jk}}^2-1} z_{\text{ik},j},z_{\text{jl}}\to z_{\text{jk}} z_{\text{kl}}+\sqrt{z_{\text{jk}}^2-1}
   \sqrt{z_{\text{kl}}^2-1} z_{\text{jl},k}.
\end{displaymath}
The jacobian for this transformation is
\begin{equation}
J(z_{ij},z_{jk},z_{kl},z_{ik,j},z_{jl,k}) =\left(z_{\text{ij}}^2-1\right) \left(z_{\text{jk}}^2-1\right)
   \left(z_{\text{kl}}^2-1\right) \sqrt{z_{\text{ik},j}^2-1}
   \sqrt{z_{\text{jl},k}^2-1}.
\end{equation}

For the $m$-th moment ($Moment_{m} \equiv \zeta_m^{'}$), 
the indicated six-dimensional integration 
of $P$ in now reparameterized form $\tilde{P}$ over the hypercube defined by $z_{ij} \in [-1,1], z_{jk} \in [-1,1], z_{kl} \in [-1,1], z_{ik,j} \in [-1,1], z_{jl,k} \in [-1,1], z_{il,jk} \in [-1,1]$ yields--including a normalization factor of $\frac{27}{32 \pi^2}$--the ("intermediate function") result 
\begin{equation}
I_{m}(\mu)= \frac{27}{32 \pi^2} \int_{-1}^{1} \int_{-1}^{1} \int_{-1}^{1} \int_{-1}^{1} \int_{-1}^{1} \int_{-1}^{1} 
\end{equation}
\begin{displaymath}
J(z_{ij},z_{jk},z_{kl},z_{ik,j},z_{jl,k}) [\tilde{P}(z_{ij},z_{jk},z_{kl},z_{ik,j},z_{jl,k}, z_{il,jk})]^m \mbox{d} z_{ij} \mbox{d}z_{jk}  \mbox{d} z_{kl} \mbox{d} z_{ik,j}
\mbox{d} z_{jl,k}  \mbox{d}  z_{il,jk}.
\end{displaymath}

\begin{figure}
\includegraphics{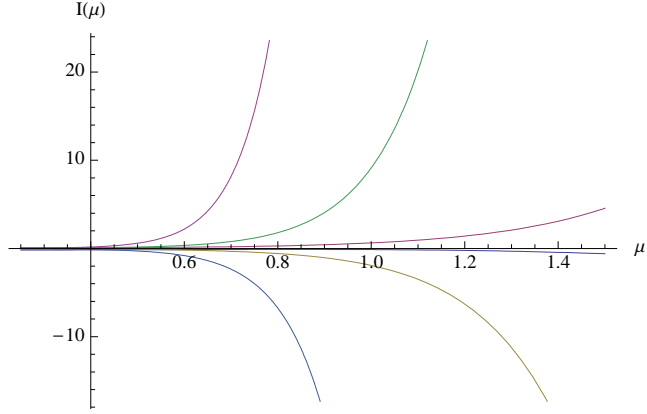}
\caption{\label{fig:intermediate}The six functions $I_{m}(\mu), m=1,...,6$. The curves for even $m$ curve upward, for odd $m$ downward, with the steepness of the curves increasing with $m$.}
\end{figure}

For the first ($m=1$) moment, we have the result 
(Fig.~\ref{fig:intermediate})
\begin{equation}
I_{1}(\mu)=-\frac{\mu ^4}{5}+\frac{34 \mu ^2}{125}-\frac{1}{5},
\end{equation}
for the second moment ($m=2$),
\begin{equation}
I_{2}(\mu)=\frac{3 \mu ^8}{35}-\frac{12 \mu ^6}{875}+\frac{20898 \mu
   ^4}{42875}-\frac{12 \mu ^2}{875}+\frac{3}{35},
\end{equation}
and for the third ($m=3$),
\begin{equation}
I_{3}(\mu)=-\frac{\mu ^{12}}{21}-\frac{54 \mu ^{10}}{875}-\frac{27873 \mu
   ^8}{42875}-\frac{466876 \mu ^6}{1157625}-\frac{27873 \mu
   ^4}{42875}-\frac{54 \mu ^2}{875}-\frac{1}{21}.
\end{equation}
At this point, we omit terms of lower order $2 j$ than $2 m$, the coefficients of which match the coefficients $C_{4m-2j}(m)$.
Then, 
\begin{equation}
I_{4}(\mu)=\frac{\mu ^{16}}{33}+\frac{584 \mu ^{14}}{5775}+\frac{278884 \mu
   ^{12}}{282975}+\frac{8984 \mu ^{10}}{4851}+\frac{65788454 \mu
   ^8}{20543985}+\dots,
\end{equation}
\begin{equation}
I_{5}(\mu)=-\frac{3 \mu ^{20}}{143}-\frac{18 \mu ^{18}}{143}-\frac{70881 \mu
   ^{16}}{49049}-\frac{2178728 \mu ^{14}}{441441}-\frac{59472398 \mu
   ^{12}}{4855851}-\frac{4103383444 \mu ^{10}}{273546273}+\ldots,
\end{equation}
\begin{equation}
I_{6}(\mu)=\frac{\mu ^{24}}{65}+\frac{2556 \mu ^{22}}{17875}+\frac{5454 \mu
   ^{20}}{2695}+\frac{3359372 \mu ^{18}}{315315} +
\end{equation}
\begin{displaymath}
+\frac{3273117 \mu
   ^{16}}{86515}+\frac{597414184 \mu ^{14}}{7872865}+\frac{173821048732
   \mu ^{12}}{1771394625} +\ldots,
\end{displaymath}
\begin{equation}
I_{7}(\mu)= -\frac{\mu ^{28}}{85}-\frac{4298 \mu ^{26}}{27625}-\frac{826637 \mu
   ^{24}}{303875}-\frac{165865636 \mu ^{22}}{8204625}-\frac{71226035 \mu
   ^{20}}{722007} -
\end{equation}
\begin{displaymath}
-\frac{1947049760374 \mu
   ^{18}}{6711055065}-\frac{93373201818911 \mu
   ^{16}}{167776376625}-\frac{33225665966177656 \mu
   ^{14}}{48487372844625} \ldots,
\end{displaymath}
\begin{equation}
I_{8}(\mu)= \frac{3 \mu ^{32}}{323}+\frac{6672 \mu ^{30}}{40375}+\frac{12986136 \mu
   ^{28}}{3674125}+\frac{4250871568 \mu
   ^{26}}{121246125}+\frac{3319251741068 \mu
   ^{24}}{14670781125}+
\end{equation}
\begin{displaymath}
 +\frac{755365923834768 \mu
   ^{22}}{826454003375}+\frac{2024301386770232 \mu
   ^{20}}{826454003375}+\frac{61510285844520752 \mu
   ^{18}}{14049718057375}+\frac{3853435310162220966 \mu
   ^{16}}{724564031244625} \ldots,
\end{displaymath}
and
\begin{equation}
I_{9}(\mu)= -\frac{\mu ^{36}}{133}-\frac{9774 \mu ^{34}}{56525}-\frac{651051 \mu
   ^{32}}{145775}-\frac{8355664 \mu ^{30}}{146965}-\frac{18384996780 \mu
   ^{28}}{39122083}-\frac{4848288282648 \mu
   ^{26}}{1944597655}-
\end{equation}
\begin{displaymath}
-\frac{133915228926036 \mu
   ^{24}}{15026436425}-\frac{61222919937476688 \mu
   ^{22}}{2809943611475}-\frac{396008663496240078 \mu
   ^{20}}{10677785723605}-\frac{2103161056387491292 \mu
   ^{18}}{47564681859695}  \ldots
\end{displaymath}
For the nine cases ($m=1,...,9$) we have been able to explicitly compute so far, the coefficients of the corresponding $4m$-degree even 
polynomials $I_{m}(\mu)$, as already indicated, are symmetric--for reasons not immediately apparent to us--around 
the $\mu^{2 m}$ term. 

The constant terms (as well as the 
coefficients of the $\mu^{4 m}$
term) are expressible as
\begin{equation} \label{coefficient0}
C_{0}(m)=C_{4m}(m)=\frac{3 (-1)^m}{4 \left(m+\frac{1}{2}\right) \left(m+\frac{3}{2}\right)}.
\end{equation}
Additionally, the coefficients of the second and $(4 m-2)$ terms are
\begin{equation} \label{coefficient2}
C_{2}(m)=C_{4m-2}(m)=\frac{3 (-1)^m m (2 m (4 m-5)-15)}{100 \left(m-\frac{1}{2}\right)
   \left(m+\frac{1}{2}\right) \left(m+\frac{3}{2}\right)}.
\end{equation}
Further, the coefficients of the fourth and $(4 m-4)$ terms are
\begin{equation} \label{coefficient4}
C_{4}(m)=C_{4m-4}(m)= \frac{3 (-1)^m m (2 m (2 m (2 m (8 m (6 m-7)+155)-13)-1017)-315)}{19600
   \left(m-\frac{3}{2}\right) \left(m-\frac{1}{2}\right)
   \left(m+\frac{1}{2}\right) \left(m+\frac{3}{2}\right)}.
\end{equation}
(These results were obtained using the  "rate", guessing program of C. Krattenthaler, based on Mathematica programming of M. Trott.) Still further, M. Trott was able to obtain the result 
(using the FindSequenceFunction command of Mathematica)
\begin{equation} \label{coefficient6}
C_{6}(m)=C_{4m-6}(m)=
\end{equation}
\begin{displaymath}
\frac{(-1)^m (m-1) m (4 m (2 m (2 m (m (4 m (20 m (4
   m-11)+173)-4303)+4733)+14911)-9165)-4725)}{529200
   \left(m-\frac{5}{2}\right) \left(m-\frac{3}{2}\right)
   \left(m-\frac{1}{2}\right) \left(m+\frac{1}{2}\right)
   \left(m+\frac{3}{2}\right)}.
\end{displaymath} 

From the formulas for these coefficients, it is 
clear that the numerator of the coefficient ($C_{2 j}(m)=C_{4 m-2j}(m)$) 
of $\mu^{2 j}$ is
a polynomial of degree  $3 j$, and the denominator
is a polynomial of degree $j+2$. (The denominators are very simple in structure (\ref{hammer})--as evidenced above.). For $j=0$, the roots are $-\frac{3}{2}$ and 
$-\frac{1}{2}$, and as $j$ increases by 1, an additional root 1 larger in value than the previous smallest is added. Thus, poles occur at the coefficient functions at such half-integers.) Utilizing this observation, we were then able--using simple fitting 
procedures--to move on to obtaining the coefficients 
$C_{8}(m)=C_{4m-8}(m)$,  
$C_{10}(m)=C_{4m-10}(m)$ $C_{12}(m)=C_{4m-12}(m)$,  
$C_{14}(m)=C_{4m-14}(m)$ and 
$C_{16}(m)=C_{4m-16}(m)$--but not yet higher. In studying the roots of these functions, we have detected one quite interesting feature. That is, as $j$ increases, the dominant roots of $C_{2j}(m)$ show very strong evidence of converging 
to $j-\frac{1}{2}$, the subdominant roots 
to $j-\frac{3}{2}$,...For instance, for $j=8$, the dominant roots of $C_{16}{m}=
C_{4m -16}(m)$ are $7.49999796, 6.4999352, 5.4980028, 4.4493216$, while for $j=7$, they are $6.5000204, 5.500556, 4.515944$. Such roots would then come increasingly close to canceling the near-to-matching poles in the denominators in $C_{2 j}(m)$ as $j$ increases.

In the second stage of our procedure to compute the $m$-th moment, we reverse the substitution (\ref{substitution}) 
in these $4 m$-degree polynomials,
multiply the result by the necessarily {\it nonnegative} factor 
$(\rho_{22} \rho_{33})^{2 m}$ (the factor $(\rho_{22} \rho_{33})$ had been removed in forming the polynomial $P$ in seven variables, proportional to $|\rho^{PT}|$) and also by the jacobian corresponding to the transformation to Bloore (correlation) variables for the real two-qubit density matrices \cite{andai}
\begin{equation}
jac= (\rho_{11} \rho_{22} \rho_{33} \rho_{44})^{\frac{3}{2}}.
\end{equation}
The result is, then, integrated over the unit three-dimensional simplex,
\begin{equation}
\rho_{11}+\rho_{22}+\rho_{33} +\rho_{44}=1, \hspace{.2in} \rho_{ii} \geq 0, \hspace{.2in} i=1,\ldots,4
\end{equation}
to obtain the $m$-th moment. In other words (taking into account the appropriate normalization factor), and setting 
$\rho_{44}=1-\rho_{11}+\rho_{22}+\rho_{33}$,
\begin{equation}
Moment_{m} \equiv \zeta_m^{'}=\frac{1146880}{\pi ^2} \int_0^1 \int_0^{1-\rho_{11}} \int_0^{1-\rho_{11}-\rho_{22}} 
\end{equation}
\begin{displaymath}
(\rho_{22} \rho_{33})^{2 m} (\rho_{11} \rho_{22} \rho_{33} \rho_{44})^{\frac{3}{2}} I_{m}(\sqrt{\frac{\rho_{11} \rho_{44}}{\rho_{22} \rho_{33}}}) \mbox{d} \rho_{33} \mbox{d} \rho_{22} 
\mbox{d} \rho_{11}. 
\end{displaymath}
We are, in fact, able to  perform the indicated symbolic integration, obtaining 
thereby
\begin{equation}
Moment_{m}=\zeta_m^{'}=\frac{1146880}{\pi ^2 \Gamma (4 m+10)} \Sigma_{i=0,2,4...}^{4 m} 
\Gamma \left(\frac{i+5}{2}\right)^2 \Gamma \left(-\frac{i}{2}+2
   m+\frac{5}{2}\right)^2 C_{i}(m)
\end{equation}
\begin{displaymath}
=\frac{2293760}{\pi ^2 \Gamma (4 m+10)} \Sigma_{i=0,2,4...}^{2 m-2} 
\Gamma \left(\frac{i+5}{2}\right)^2 \Gamma \left(-\frac{i}{2}+2
   m+\frac{5}{2}\right)^2 C_{i}(m) +
\end{displaymath}
\begin{displaymath}
+\frac{1146880}{\pi ^2 \Gamma (4 m+10)} \Gamma \left(m+\frac{5}{2}\right)^4 C_{2 m}(m),
\end{displaymath}
where the  $C_{i}(m)$'s  are our previously-indicated rational functions 
((\ref{coefficient0})-(\ref{coefficient6})), symmetric about
$2 m$. These (rational functions) $C_{i}(m)$'s 
themselves are the ratios of polynomials in $m$ of degree $\frac{3 i}{2}$ divided by 
 the term (using the Pochhammer symbol, as well as rising factorials for gamma functions with half-integer arguments)
\begin{equation} \label{hammer}
\mbox{denominator}(C_{i}(m)) 
=\left(\frac{1-i}{2}+m\right)_{\frac{i}{2}+2} = 
\frac{\Gamma \left(m+\frac{5}{2}\right)}{\Gamma
   \left(-\frac{i}{2}+m+\frac{1}{2}\right)}=
\frac{2^{\frac{i}{2}+2} (2 m+3)\text{!!}}{(-i+2 m-1)\text{!!}}
\end{equation}
\begin{displaymath}
=\Pi_{k=-2,0,...}^{i} (m+\frac{1-k}{2}).
\end{displaymath}
For $i=4$, by way of example, this gives us (\ref{coefficient4})
\begin{equation}
\left(m-\frac{3}{2}\right) \left(m-\frac{1}{2}\right)
   \left(m+\frac{1}{2}\right) \left(m+\frac{3}{2}\right).
\end{equation}
On the other hand, the {\it numerators} of the $C_{i}(m)$'s for $m>0$
have zero as a trivial root, and for $m> 4 n$, 
trivial roots $0, \ldots n$.

Again, converting gamma functions with half-integer arguments to rising factorials, we have, equivalently, that 
\begin{equation}
Moment_m=\zeta_m^{'}=
\end{equation}
\begin{displaymath}
35 \frac{ 2^{7-4m}}{ \Gamma[4 m+10]} \Big[ \Big((2m+3)!!\Big)^2 C_{2 m}(m) +2 \Sigma_{i=0,2,4...}^{2 m-2}
\Big((3+i)!! (3-i+4 m)!!\Big)^2 C_i(m) \Big].
\end{displaymath}

Pursuant to these formulas, the first moment (mean) of the Hilbert-Schmidt probability distribution of $|\rho^{PT}|$
over the interval $[-\frac{1}{16},\frac{1}{256}]$ is (departing from the convention of denoting moments by
$\mu$, since that symbol has been employed in our earlier studies \cite{slaterPRA2,advances} 
and above (\ref{substitution}))
\begin{equation} \label{firstmoment}
\zeta_1^{'}=-\frac{1}{858} =-\frac{1}{2 \cdot 3 \cdot 11 \cdot 13} 
\approx -0.0011655,
\end{equation}
falling within the [negative] region of entanglement. Then, successively,  the ([necessarily] decreasing in absolute value) raw (non-central) moments are
\begin{equation}
\zeta_2^{'}=\frac{27}{2489344} =\frac{3^3}{2^{10} \cdot 11 \cdot 13 \cdot 17}\approx 0.0000108462,
\end{equation}
\begin{equation}
\zeta_3^{'}=-\frac{8363}{66216550400} = -\frac{8363}{2^{13} \cdot 5^2 \cdot 7 \cdot 11 \cdot 13 \cdot 17 \cdot 19} \approx -1.2629773 \cdot 10^{-7},
\end{equation}
\begin{equation}
\zeta_4^{'}=\frac{21859}{10443295948800} = 
\frac{21859}{2^{17} \cdot 3 \cdot 5^{2} \cdot 11 \cdot 13 \cdot 17 \cdot 19 \cdot 23} \approx 2.09311 \cdot 10^{-9},
\end{equation}
\begin{equation}
\zeta_5^{'}= -\frac{23071}{539633583390720} =- \frac{23071}{2^{18} \cdot 3 \cdot 5 \cdot 7^2 \cdot 13 \cdot 17 \cdot 19 \cdot 23 \cdot 29}\approx-4.27531 \cdot 10^{-11},
\end{equation}
\begin{equation}
\zeta_6^{'}=\frac{3317321}{3253917653076541440} = 
\frac{7 \cdot 43 \cdot 103 \cdot 107}{2^{28} \cdot 3 \cdot 5 \cdot 11^2 \cdot 17 \cdot 19 \cdot 23 \cdot 29 \cdot 31} \approx 1.01949 
\cdot 10^{-12},
\end{equation}
\begin{equation}
\zeta_7^{'}=-\frac{419856257}{15366774022001834065920} =
\end{equation}
\begin{displaymath}
-\frac{43 \cdot 2179 \cdot 4481}{2^{30} \cdot 3^4 \cdot 5 \cdot 11 \cdot 13 \cdot 17 \cdot 19 \cdot 23 \cdot 29 \cdot 31 \cdot 37}
\approx -2.73223 \cdot 10^{-14},
\end{displaymath}
\begin{equation} 
\zeta_8^{'}=\frac{16945249}{21117403549591928832000} =
\end{equation}
\begin{displaymath}
\frac{109 \cdot 155461}{2^{33} \cdot 3 \cdot 5^3 \cdot 11  \cdot 19 \cdot 23 \cdot 29 \cdot 31 \cdot 37 \cdot 41}
\approx 8.02431  \cdot 10^{-16},
\end{displaymath}
and (requiring four days of computation on a MacMini machine)
\begin{equation} \label{lastmoment}
\zeta_9^{'}=-\frac{6102620963}{240565904621616585139814400} =
\end{equation}
\begin{displaymath}
-\frac{19 \cdot 199 \cdot 1614023}{2^{37} \cdot 3 \cdot 5^2 \cdot 11^3  \cdot 13  \cdot 23 \cdot 29 \cdot 31 \cdot 37 \cdot 41 \cdot 43}
\approx -2.53678  \cdot 10^{-17}.
\end{displaymath}
(After four weeks of uninterrupted computation, we did not succeed, however, in determining $\zeta_{10}^{'}$.)

Interestingly, the sequence of denominators  immediately above 
(in apparent contrast to that of the numerators) appears to 
be "nice" in that the number of their prime factors do not grow rapidly, but rather linearly. This is a strong indication of the possible existence of 
a "closed form", that is 
an expression which is built by forming products and quotients of factorials \cite[fn. 12]{determinantcalculus}.

The skewness ($\gamma_1$) of the Hilbert-Schmidt probability distribution over $|\rho^{PT}|$ is negative (as well as all moments listed of odd order), that is, -3.13228--so, the left tail of the distribution is more pronounced than the right tail--while its kurtosis ($\gamma_2$), a measure of "peakedness" is quite high, 17.6316.
(Higher kurtosis indicates that more of the variance is the result of infrequent extreme deviations than frequent modestly sized 
deviations.)
From the first two moments, we obtain the variance
\begin{equation}
\sigma^2=\frac{30397}{3203785728} \approx 9.487838 \cdot 10^{-6}.
\end{equation}

Application of the standard-form one-sided Chebyshev inequality \cite{marshall} 
(we perform a linear transformation,  so that negative values of
$|\rho^{PT}|$ are mapped to [0,1]), then, yields an upper bound on the Hilbert-Schmidt separability probability of the real two-qubit density matrices of 
$\frac{30397}{34749} \approx 0.874759$. This is a substantially weaker upper bound than that of $\frac{1129}{2100} \approx 0.537619$ established in \cite{advances}, by enforcing the nonnegativity of pairs of $3 \times 3$ principal minors of the partial transpose, as well as weaker than 
$\frac{1024}{135 \pi^2} \approx 0.76854$, obtained by requiring the nonnegativity
of all six $2 \times 2$ principal minors of the partial transpose 
\cite{advances}.

Using the general formulae for the coefficients ((\ref{coefficient0})-(\ref{coefficient4}))--derived above, using the "rate" program of C. Krattenthaler--of
the zero-th, second, fourth (and by symmetry) the $4m, 4m-2$ and $4m-4$ powers of $\mu$ in the intermediate functions $I_{m}(\mu)$, we have been able to obtain the exact contribution of the associated six terms to the 
$m$-th moment. This contribution is 
the product of the two factors
\begin{equation}
\frac{945 (-1)^m (m (2 m (2 m (2 m (40 m (6 m-5)-169)+101)-495)-9)+27)}{2
   \left(16 m^4-40 m^2+9\right)}
\end{equation}
and
\begin{equation}
\frac{256 \Gamma \left(2 m+\frac{1}{2}\right)^2}{\pi  \Gamma (4
   m+10)}+\frac{2^{-8 m} \Gamma (4 m+8)}{(m+2) (4 m+1)^2 (4 m+3)^2 (4
   m+5)^2 (4 m+7)^2 (4 m+9) \Gamma (2 m+4)^2}.
\end{equation}
For $m=3$, the ratio of the true/known moment to the product of these two
factors is 1.05766, increasing monotonically, in a quasi-linear manner,  to 1.94638 for $m=9$. Extending this form of analysis/approximation to take into account the exact formulas we have also so far obtained 
for $C_{2j}(m)=C_{4m-2j}(m)$, for $j=3,\ldots,8$, we can reduce this ratio from 1.94638 to $\frac{170368623463798669312}{164584930558733068259} 
\approx 1.03514$.

\section{Hilbert-Schmidt probability distribution over $|\rho|$} 
\label{secnotranspose}
The determinant of a $4 \times 4$ density matrix itself is restricted to a smaller   range $[0,\frac{1}{256}]$ than that--$[-\frac{1}{16},\frac{1}{256}]$--of its partial transpose. We have computed the initial moments of the Hilbert-Schmidt probability distribution of $|\rho|$ over this interval, where $\rho$ corresponds to a generic real two-qubit system. (Doing so involves only a series of {\it three}-fold integrations \cite{szHS,tbs}--three being the number of independent eigenvalues of a $4 \times 4$ density matrix--rather than the {\it nine}-fold [6+3] integrations used above for the moments of the probability distribution of $|\rho^{PT}|$.) The first moments 
($\tilde{\zeta}_i^{'}$) are 
\begin{equation} \label{firstmomentB}
\tilde{\zeta}_1^{'}=\frac{1}{2288} = (2^4 \cdot 11 \cdot 13)^{-1} \approx 0.000437063,
\end{equation}
\begin{equation}
\tilde{\zeta}_2^{'}=\frac{1}{2489344} 
= (2^{10} \cdot 11 \cdot 13 \cdot 17)^{-1} \approx 4.01712 \cdot 10^{-7},
\end{equation}
\begin{equation}
\tilde{\zeta}_3^{'}=-\frac{1}{1891901440} = (2^{13} \cdot 5 \cdot 11 \cdot 13 \cdot 17 \cdot 19)^{-1} \approx 5.28569 \cdot 10^{-10},
\end{equation}
\begin{equation}
\tilde{\zeta}_4^{'}=\frac{3}{3481098649600} =
(2^{17} \cdot 5^2  \cdot 11 \cdot 13 \cdot 17 \cdot 19 \cdot 23)^{-1}\approx 8.61797 \cdot 10^{-13},
\end{equation}
\begin{equation}
\tilde{\zeta}_5^{'}= 
 \frac{1}{616724095303680} =
(2^{21} \cdot 3  \cdot 5  \cdot 7  \cdot 13 \cdot 17 \cdot 19 \cdot 23 \cdot 29)^{-1}  
\approx 1.62147 \cdot 10^{-15},
\end{equation}
\begin{equation}
\tilde{\zeta}_6^{'}=\frac{1}{295810695734231040} \approx 3.38054 \cdot 10^{-18},
\end{equation}
\begin{equation}
\tilde{\zeta}_7^{'}=\frac{1}{131339948905998581760}
\approx 7.61383 \cdot 10^{-21},
\end{equation}
\begin{equation}
\tilde{\zeta}_8^{'}=\frac{1}{54905249228939014963200} \approx 1.82132 \cdot 10^{-23},
\end{equation}
\begin{equation}
\tilde{\zeta}_9^{'}=\frac{1}{21869627692874235012710400} \approx 4.57255 
\cdot 10^{-26}.
\end{equation}
\begin{equation}
\tilde{\zeta}_{10}^{'}= \frac{1}{8372488740021088953229639680} \approx 1.19439 
\cdot 10^{-28}
\end{equation}
\begin{equation}
\tilde{\zeta}_{11}^{'}= \frac{1}{3100931560849549878551107338240} \approx 3.22484 
\cdot 10^{-31}
\end{equation}
\begin{equation}
\tilde{\zeta}_{12}^{'}= \frac{1}{1116717015021019439340374162669568} \approx 8.95482 
\cdot 10^{-34}
\end{equation}
\begin{equation}
\tilde{\zeta}_{13}^{'}= \frac{3}{1177747849688102259981247358666014720} \approx 2.54723  
\cdot 10^{-36}
\end{equation}
\begin{equation}
\tilde{\zeta}_{14}^{'}=\frac{1}{135156718750413942110951421022086103040}
\approx 7.39882 \cdot 10^{-39}
\end{equation}
and
\begin{equation} \label{lastmomentB}
\tilde{\zeta}_{15}^{'}=\frac{1}{45686458852962503761039927438910834081792}
\approx 2.18883 \cdot 10^{-41}.
\end{equation}
\section{Relations between the two sets of moments}
Similarly to  the sequence ((\ref{firstmoment})-(\ref{lastmoment})) of denominators of the earlier set of moments presented, the prime factors of the denominators in this latter set of  moments (and nine more we have been able to compute) do not grow rapidly, indicative of the possibility of a closed form for them. (Of course, the numerators here--mostly 1's with two 3's--unlike
the earlier ones, are comparatively well-behaved.) For instance, 
\begin{equation}
\tilde{\zeta}_{24}^{'}= (2^{98} \cdot 3^2 \cdot 5^2 \cdot 11 \cdot 13^2 \cdot 17 \cdot 19 \cdot 29 \cdot 31 \cdot 53 \cdot 59 \cdot 61 \cdot 67 \cdot 71 \cdot 73 \cdot 79 \cdot 83 \cdot 89 \cdot 97 \cdot 101 \cdot 103)^{-1} \approx 6.66035 \cdot 10^{-64}.
\end{equation}

Interestingly--but for the cases $m=1,5$--the powers to which 2 is raised in the denominators of the $m$-th entries ($m=1,...,9$) of the two moment sequences 
((\ref{firstmoment})-(\ref{lastmoment}) and (\ref{firstmomentB})-(\ref{lastmomentB})) coincide. (In our two sequences, none of the powers of 2 occurring in the denominators is a  square number.)
More strikingly still, taking 11 to be the fifth prime number, the highest prime occurring in the denominator of the $m$-th member of each of the two sequences is the same (5+m)-th prime.

\begin{figure}
\includegraphics{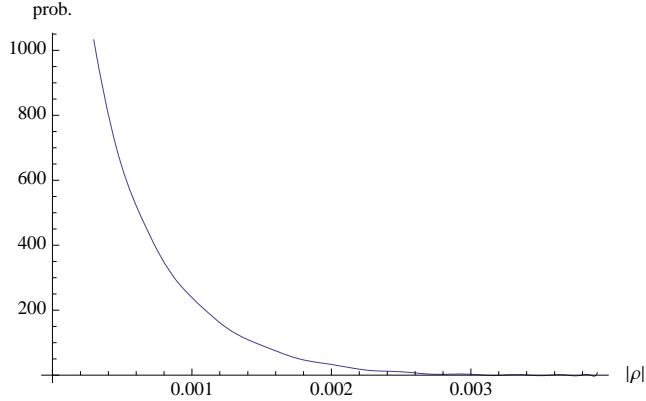}
\caption{\label{fig:NoTransposeFit}Fit of a twenty-four degree polynomial
to the first twenty-four moments of the Hilbert-Schmidt probability distribution over $|\rho|$, where $\rho$ is a generic real two-qubit density matrix}
\end{figure}

\begin{figure}
\includegraphics{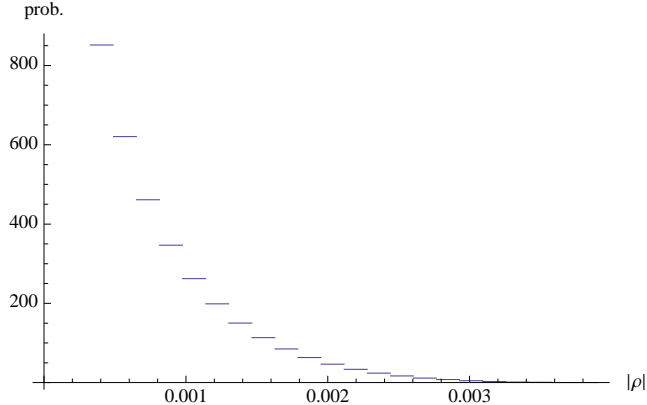}
\caption{\label{fig:NoTransposeFitMnat}Reconstruction of the Hilbert-Schmidt probability distribution over $|\rho|$, based on its first twenty-four moments, using the stable approximation method advanced in 
\cite[eq. (6)]{mnatsakanov}}
\end{figure}

In Fig.~\ref{fig:NoTransposeFit}, we display the fit of a power series in 
$|\rho|$ of degree twenty-four (with twenty-five unknown coefficients) to the computed first twenty-four moments.
(We also--giving us the needed twenty-fifth constraint--require the "zeroth" moment to be 1, as mandated for any probability distribution.) No nonnegativity constraints were, however, imposed and certain slight
incursions into negative regions result. (The "probability" mass below the  $|\rho|$ axis is 0.000397, while above it, the mass is 1.000397.) The distribution is clearly peaked at $|\rho|=0$, the locus of the degenerate
(pure, pseudo-pure,...) states, those having at least one eigenvalue zero.
In Fig.~\ref{fig:NoTransposeFitMnat}, we attempt an alternative reconstruction of this probability distribution using the stable approximation method advanced in \cite[eq. (6)]{mnatsakanov}, giving us a sequence of plateaus.

Further, inputting the first twenty-four moments computed into the FindSequenceFunction command of Mathematica, we obtained the formula 
(again employing the Pochhammer symbol)
\begin{equation} \label{generalformula}
\tilde{\zeta}_{m}^{'}=\frac{2^{1-8 m} (1)_m \left(\frac{3}{2}\right)_m}{(m+2)
   \left(\frac{11}{4}\right)_m \left(\frac{13}{4}\right)_m}=
\frac{945 \sqrt{\pi } 2^{-8 m-4} \Gamma (2 m+2)}{(m+2) \Gamma \left(2
   m+\frac{11}{2}\right)}.
\end{equation}
The associated {\it moment-generating function} is the generalized 
hypergeometric function
\begin{equation}
M(t)=\frac{4032 \left(\,_3F_3
\left(\frac{1}{2},1,1;\frac{7}{4},2,\frac{9}{4};
\frac{t}{256}\right)-1\right)}{t}.
\end{equation}
The inverse Fourier transform of the associated characteristic function, that is $M(i t)$, then, should yield the Hilbert-Schmidt probability distribution over $|\rho|$ for generic real two-qubit density matrices $\rho $. (However, we have 
not been able to explicitly evaluate it.)

\begin{figure}
\includegraphics{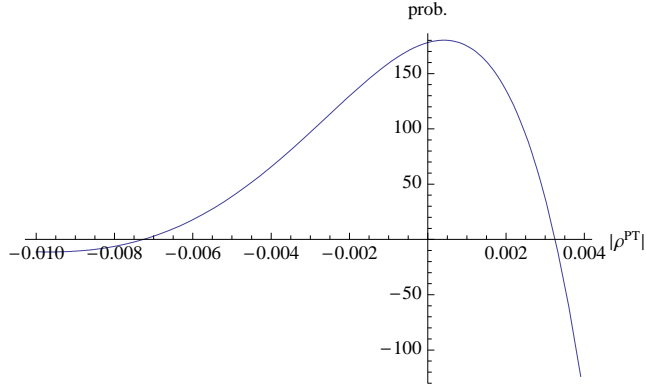}
\caption{\label{fig:TransposeFit}Fit of a nine-degree polynomial
to the first nine moments of the Hilbert-Schmidt probability distribution over $|\rho^{PT}|$, where $\rho$ is a generic real two-qubit density matrix.  The domain of separability is $|\rho^{PT}| >0$}
\end{figure}

In Fig.~\ref{fig:TransposeFit}, we display the fit of a power series in 
$|\rho|^{PT}$ of degree nine to the computed first nine moments 
((\ref{firstmoment})-(\ref{lastmoment})) (cf. \cite[Figs. 1, 2]{giraud1}) 
of the Hilbert-Schmidt probability distribution over $|\rho^{PT}|$, where  $\rho$ is a generic real two-qubit density matrix.
No nonnegativity constraints were, however, imposed and considerable
incursions into negative regions now result. (Such negativity can be obviated through the use of maximum-entropy, spline-fitting and other methodologies 
\cite{john,mnatsakanov}, and we intend to explore such directions. The use of the stable approximation approach \cite{mnatsakanov} used to produce
Fig.~\ref{fig:NoTransposeFitMnat} was not insightful with our relatively small number of explicit moments.)

Since the plotted distribution (Fig.~\ref{fig:TransposeFit}) appears to be {\it unimodal}, one can presumably use the computations of the first and second moments above to isolate the mode of the distribution within the interval \cite[eq. (13)]{gavriliadis1}
\begin{equation}
\left\{-\frac{1}{858}-\frac{\sqrt{\frac{30397}{51}}}{4576},
\frac{\sqrt{\frac{30397}{51}}}{4576}-\frac{1}{858}\right\} =
\{-0.00650062, 0.00416962\},
\end{equation}
containing $|\rho^{PT}|=0$. Narrower intervals containing the mode can be obtained using
higher-order moments and the associated Hankel determinants 
\cite[Thm. 3.2]{gavriliadis1}.
\section{Concluding Remarks}
It is clear that it would be of considerable utility  to have available
exact values for still higher-order (than $m=9$) moments--and for the coefficients $C_{2j}(m)$ of the terms in the intermediate functions 
$I_{m}(\mu)$--but the associated computational demands seem quite considerable. 
(The Hilbert-Schmidt separability probability predicted by the curve in 
Fig.~\ref{fig:TransposeFit}--that is the "probability mass" lying within the interval $[0,\frac{1}{256}]$--is $0.39648$, while our previous studies 
\cite{advances}, indicate that the actual value is considerably higher, $\approx 0.45$--a discrepancy still higher-order moments should ameliorate.) Also, of course, it would be interesting to extend our forms of analyses from the {\it real} case to the more fully general setting of the 15-dimensional convex set of {\it complex} two-qubit ($4 \times 4$) density matrices. But, at this stage of development of our technical apparatus, we are unable even to  compute the corresponding first Hilbert-Schmidt moment (mean) over $|\rho^{PT}|$ (known to be $-\frac{1}{858}$ in the real two-qubit case). However, progress, in these regards,  should be much more readily achievable in terms of the moments over $|\rho|$.

In fact, inputting the first twenty such moments computed into the FindSequenceFunction command of Mathematica, we obtained the formula 
(again employing the Pochhammer symbol)
\begin{equation} \label{generalformula2}
\tilde{\zeta}_{m/complex}^{'}=\frac{256^{-m} (1)_m (2)_m (3)_m}{\left(\frac{17}{4}\right)_m
   \left(\frac{9}{2}\right)_m \left(\frac{19}{4}\right)_m}=
\frac{108972864000 \Gamma (m+1) \Gamma (m+2) \Gamma (m+3) \Gamma
   (m+4)}{\Gamma (4 m+16)}.
\end{equation}
The associated {\it moment-generating function} is the generalized 
hypergeometric function
\begin{equation}
M(t)_{complex}=\,_3F_3\left(1,2,3;\frac{17}{4},\frac{9}{2},\frac{19}{4};\frac{t}{256}\right).
\end{equation}
Similarly, for the quaternionic two-qubit case,
\begin{equation}
\tilde{\zeta}_{m/quat}^{'}=
\frac{315071454005160652800000 \Gamma (m+1) \Gamma (m+3) \Gamma (m+5)
   \Gamma (m+7)}{\Gamma (4 (m+7))},
\end{equation}
and
\begin{equation}
M(t)_{quat}=\,_3F_3\left(1,3,5;\frac{29}{4},\frac{15}{2},\frac{31}{4};\frac{t}{256}\right)
\end{equation}

We have also been able to compute exactly the first twelve moments of the 
probability distribution of $|\rho|$ over $[0,6^{-6}]$, where $\rho$ is a generic
complex qubit-{\it qutrit} ($6 \times 6$) density matrix, but this seemed to be an insufficient number of moments to discern a general formula. 
All eleven moments found were the reciprocals of positive integers.
The first moment (mean) was $\frac{1}{4496388} =(2^2 \cdot 3 \cdot 13 \cdot 19 \cdot 37 \cdot 41)^{-1}$, while the second moment was smaller than this by a factor of $\frac{1}{1533939}$. (The twelfth moment is approximately
$6.16876 \cdot 10^{-68}$.)

\begin{acknowledgments}
I would like to express appreciation to the Kavli Institute for Theoretical
Physics (KITP)
for computational support in this research, to Michael Trott
for lending his Mathematica expertise, and Christian Krattehthaler and Mihai Putinar for general discussions and insights.
\end{acknowledgments}

\bibliography{Moments6}% Produces the bibliography via BibTeX.

\end{document}